\documentclass[12pt,preprint]{aastex}
\usepackage{latexsym,amsfonts,amsbsy,amssymb}
\usepackage{amsmath,amsthm}
\tighten
\def\spose#1{\hbox to 0pt{#1\hss}}

\def\beq{\begin{equation}}
\def\enq{\end{equation}}

\def\lta{\mathrel{\spose{\lower 3pt\hbox{$\mathchar"218$}}
     \raise 2.0pt\hbox{$\mathchar"13C$}}}
\def\gta{\mathrel{\spose{\lower 3pt\hbox{$\mathchar"218$}}
     \raise 2.0pt\hbox{$\mathchar"13E$}}}
\begin{document}

\title{A Morphological Approach to the Pulsed Emission from Soft Gamma Repeaters}

\author{ J. J. Jia\altaffilmark{1,3}, Y. F. Huang\altaffilmark{1,2} and K. S. 
Cheng\altaffilmark{1}}

\affil{$^{1}$ Department of Physics,  University of Hong Kong,
       Pokfulam Road, Hong Kong, China\\
       $^{2}$ Department of Astronomy, Nanjing University, Nanjing 210093, China\\
 $^{3}$ Department of
Physics and Astronomy, Johns Hopkins University, Baltimore, MD
21218, USA }

\begin{abstract}

We present a geometrical methodology to interpret the periodical
light curves of Soft Gamma Repeaters based on the magnetar model
and the numerical arithmetic of the three-dimensional
magnetosphere model for the young pulsars. The
hot plasma released by the star quake is trapped in the
magnetosphere and photons are emitted tangent to the local magnetic field lines.
The variety of radiation morphologies in the burst tails and the
persistent stages could be well explained by the trapped fireballs
on different sites inside the closed field lines. Furthermore, our
numerical results suggests that the pulse profile evolution of SGR
1806-20 during the 27 December 2004 giant flare is due to a
lateral drift of the emitting region in the magnetosphere.

\end{abstract}


\section{Introduction}

Soft Gamma Repeaters (SGRs) seemed weird since the first discovery
on 1979 March 5 \citep{mazets79} for their mysterious
characteristics such as the large energy release, the repetitive
emission of bursts in hard X-rays or soft gamma-rays bands, and
the pulsed periodical emissions after the bursts and in the
quiescent stages whose morphologies are both energy-dependent and
time-dependent (see \cite{woods04} for a recent review). So far,
the catalogue\footnote{http://www.physics.mcgill.ca/$\sim$
pulsar/magnetar/main.html} has four SGRs confirmed plus one
candidate. SGRs are found to be associated with young ($\sim 10^4$
yr) supernova remnants (SNRs), and their spin periods are $5\sim
8\rm{s}$ and at a large spin down rate about
$10^{-11}\rm{s~s^{-1}}$, which give an inferred ultra-strong
dipolar magnetic field at the order of $10^{14} \sim
10^{15}\rm{G}$.

A variety of models were proposed to understand the physics of
SGRs, and the magnetar model is now widely accepted to ascribe
SGRs as neutron stars with magnetic field of $10^{14}\sim
10^{15}$G \citep{td95,td96}. Unlike most of the pulsars in the
neutron star family powered by their spin-down, the
high-luminosity bursts and the persistent X-ray or soft
$\gamma$-ray pulsations of magnetars come from the decay of their
ultra-strong magnetic fields. The spectrum of the persistent X-ray
emission could be fitted by a superposition of a blackbody
component and a power law, which suggests that besides the
radiation from the neutron star surface, there is another
component coming from the magnetosphere.

The high-luminous burst has now been successfully interpreted by
the dissipation of magnetic energy. However, the persistent
long-period pulsations in the quiescent X-ray emission of SGRs are
still not well understood for their complicated and astonishing
morphologies. \cite{tlk02} assumed that the angular pattern of the
X-ray flux are modified by the resonant cyclotron scattering at a
distance about 50-100 km above the neutron star, and multi-pulses
are generated by some effects associated with the twisted
magnetosphere (e.g., the optical depth is thin near two magnetic
poles, and thick at the magnetic equator). However, their
numerical results from the Monte-Carlo simulation has a bias in
favor of the orthogonal dipole (i.e., the magnetic axis is
perpendicular to the spin axis), and the viewing angel has also to
be nearly $90^0$ from the spin axis, for the photons escaping from
optically thin poles \citep{thompson07}. In addition, the
separation of the strong peaks in their calculations is always
$\pi$ (half phase cycle), which differs from the observation.
Thompson (private communication, 2005) agreed that the change in
the persistent pulse profile reflects a re-distribution of
persistent currents on closed field lines. However, he ascribed
such re-distribution as the gradual change in the magnetic twist
at a distance of 30-100 km, while our understanding for that is
due to the gradual migration of the emitting regions (crustal
platelet motion or hot plasma drifting in the magnetosphere, we
will discuss the possibility of each candidate in \S\ref{app}).

In this paper, we present some simulated pulsed profiles by
assuming the radiation region located in the closed field lines,
and make attempts to simulate the radiation morphology evolution
in one particular event, i.e., SGR 1806-20 burst on 27 December
2004. We first have a general review on the timing properties of
SGRs in \S\ref{obs}. In \S\ref{model}, we present the motivation
to reproduce the light curves in the closed field lines and the
resultant profiles. In \S\ref{app}, we applied the geometrical
model to the well-known December 2004 burst of SGR 1806-20 and
calculate the radiation morphology by a three-dimensional
magnetosphere simulation. We also try to explain the change in the
persistent pulse profile. A brief discussion is given in
\S\ref{dis}.

\section{Periodical Emissions from SGRs\label{obs}}

The periodical emissions of the four confirmed SGRs have been
detected in the decay of burst and the quiescent or persistent
stages, with the periods in the 5-8 seconds range which are the
spin periods of the neutron stars. The pulse profiles show many
interesting and even surprising morphologies, some of which are
totally different from the canonical radio or high energy pulsars.
The characteristics of light curves of the X-ray and $\gamma$-ray
pulsars such as number of peaks, the peak separation and the
relative amplitudes of the peaks are unchangeable, while those of
the SGRs are time dependent. Here, we list several main features
of the persistent emissions from SGRs:

1. Multi-peaked morphology: The most dramatic example is the pulse
profile change of SGR 1900+14 after the outburst on 1998 August
27. A four-peaked repetitive pattern of the X-ray light curve was
detected by both \textit{Ulysses} and \textit{BeppoSAX} half a
minute after the burst onset \citep{hurley99,feroci01}, and these
peaks were found to be evenly spaced at 1.0 s intervals on the
5.16 s rotation period. And several minutes later, this
multi-peaked profile evolved as a simple sinusoidal
morphology\citep{woods04}, and this evolution in pulse profile
lasted for a couple of years. Such change in morphology is also
found in the \textit{RXTE} PCA archive of SGR 1806-20 between 1996
and 2005\citep{woods07}. During the 10-year observation, the 2-10
keV pulse shape evolved from one broad peak pattern to a
three-peaked one in 2003, and then the sinusoidal shape again
until the multi-peaked profile after the burst on December 2004
(see Fig. 3 in \cite{woods07}).

2. Relative magnitudes of peaks: Besides the evolution of the
number of peaks, the relative magnitudes of the peaks in one phase
cycle also change with time. \cite{palmer05} showed such pulse
profile evolution during the giant flare of 27 December 2004. The
folded light curves in different time intervals from 30 to 265
seconds following the main spike indicate the growth of the second
peak, whose intensity related to the primary peak increases from
the DC level to nearly equal in height. In the other words, we can
say that the primary peak fades until the same magnitude as the
secondary one. At the late stage of the decay of the giant flare,
the relative magnitude of the third peak 0.2 in phase prior to the primary
one starts to grow up.

3. Energy-dependent profiles: The evolution of pulse profiles of
SGRs is not only time-dependent, but also changes in different
energy bands. \cite{woods07} investigated the energy dependence of
the SGR 1806-20 pulse profiles in three energy bands between 2 to
40 keV. Six months before the flare, there was only one broad peak
in the pulse profile below 15 keV, and it showed two clear peaks
in the 15-40 keV band. After the giant flare, the light curve
became more complicated, showing multiple peaks in all energy
intervals, and the peaks were inconsistent in phase (see Fig. 4 in
\cite{woods07}).

The totally different phenomena require the totally different
physics in this small (may be not small) group of neutron stars,
compared with the canonical pulsars. The thermal spectrum
component suggests that the radiation partly comes from the hot
spots on the stellar surface. In addition, the phase inconsistence
of the pulse profile indicates that it may not be localized in a
particular region, and the non-predictable star quakes can produce
the randomly-localized emitting regions during every burst event.
This is the main motivation for us to build up the model of the
alterable pulse profiles in the next section.

\section{Radiations from the Closed Field lines\label{model}}

\subsection{Why closed regions?\label{reason}}

The theoretical models for radiation from high energy pulsars
(e.g. Crab and Vela) require that the radiation engine is located
in the open field line region no matter in the polar gap model
\citep{harding81,dh96}, or the outer gap model
\citep{chr86,cr94,romani95,crz00}, or the modified outer gap model
\citep{dr03,jia07}. However, we cannot apply them to magnetars to
explain their persistent X-ray emissions. We assume that the
pulsed emissions of SGRs in the decay or afterglow of the bursts
are originated in the closed regions rather than in the open
regions. We have three main reasons for confining the radiation
regions to the closed zones in the neutron star magnetosphere:

1. The volume occupied by the open field lines is much smaller
compared to that by the closed field lines. Take SGR 1806-20 for
example, the spin period is $P=7.56~\rm{s}$, which means the
radius of the light cylinder reaches
$R_L=\frac{cP}{2\pi}=3.6\times 10^{10}~\rm{cm}$, where $c$ is the
speed of light. Thus, the radius of the polar cap is
$R_p=R_{\ast}\sqrt {R_{\ast}/R_L}=5.3\times 10^{3}~\rm{cm}$,
    corresponding to an angular size of $\theta_p=0.3^0$,
much smaller than the characteristic size of the crustal platelet, where
$R_{\ast}=10^{6}~\rm{cm}$ is the stellar radius. So, the polar cap
area is only $10^{-5}$ of the neutron star surface, and there is
no reason why the crustal platelet, where the magnetic energy is released,
should be located in the open area.

\begin{figure}
\epsscale{1}
\plotone{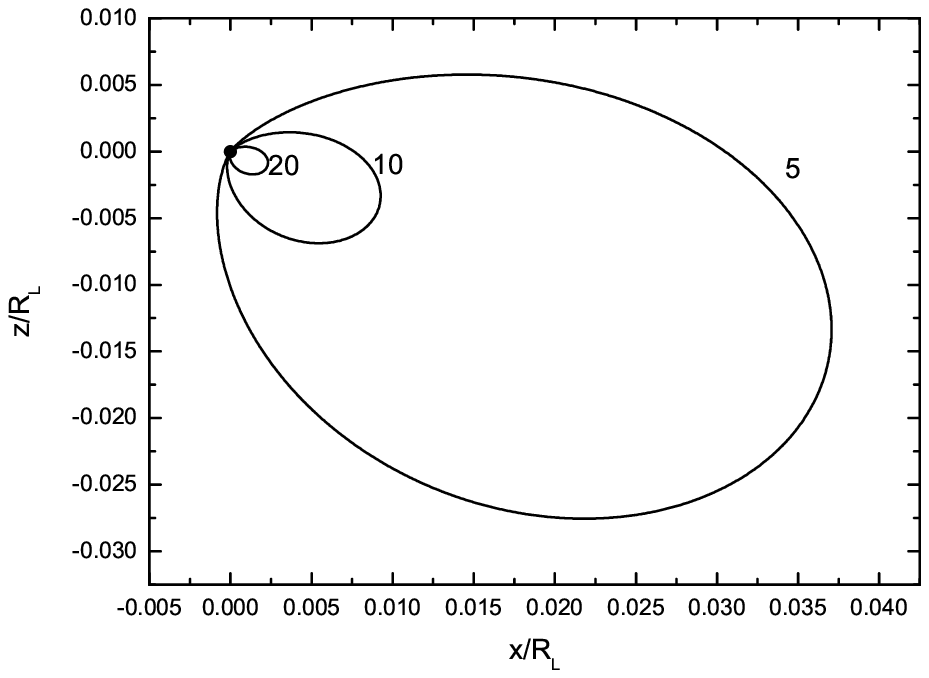}

\caption{The closed field lines on the
\mbox{\boldmath$\Omega$}-\mbox{\boldmath$\mu$} plane of layer
parameters $a_1=5, 10, 20$. The inclination angle is
$\alpha=30^0$, and spin period of SGR 1806-20 (P=7.56 s) is
applied. The neutron star is located at the point (0,0).
\label{fig1}}
\end{figure}

2. The open field lines reach the light cylinder, where the
co-rotating speed approaches the speed of light, and the
relativistic effects play a significant role in the radiation
morphology \citep{romani95}. Such effects lead to the sharp and
narrow peaks of the light curves, and both peaks are produced by
one single pole.
However, those effects become less important when it is applied to
the closed field lines, for they are much closer to the stellar
surface than the open lines.
As shown in Fig.\ref{fig1}, we find that the farthest distance
where the closed field lines can reach drops much quickly when
their footprints are displaced further away from the polar cap.
For the closed line on the plane (on which both the magnetic axis
and spin axis lie) with layer parameter $a_1=5$, it only reaches a
distance less than $0.05R_L$ away from the stellar surface, and
$0.01R_L$ for $a_1=10$ (the definition of $a_1$ is given in
\S\ref{stra}). Thus, the double-peaked profiles are not necessary
for closed field lines, and the broad peak could be a general
product instead of the sharp one. Other features, like the peak
separation and the multi-peaks, could also be explained by the
closed field lines (details to be discussed in the next section).

3. The emitting region has not to be localized on some particular
sites on neutron star surface or inside the magnetosphere while
in the open field lines, the accelerating gap is restricted to the
null charge surface \citep{gj69,crz00}. This freedom in loci
ensures the variety of the pulse profiles during different
outbursts, for we can argue that time-dependent light curves
result from different bunches of magnetic field lines where the
plasma is trapped.

\subsection{Strategies for numerical simulation\label{stra}}

Since we argue that the periodic pulsed radiation of magnetars in
the quiescent state is released from the closed field line regions
rather than the open one, we could apply the 3D magnetosphere
model \citep{crz00} to simulate the pulse profiles. The boundary
of these two different regions is defined as a bunch of so called
the last closed or the first open magnetic field lines, which are
tangential to the light cylinder. The hot plasma is trapped in the
closed region and oscillates along the closed field lines, and
photons are assumed to be emitted outwardly along the tangential direction
of the magnetic field lines, which makes the pulsed radiation much
different from the one generated in the open area. In order to be
consistent with the pulsar magnetosphere calculation in the
high-energy pulsar models, we adopted the same definition for the
coordinates of footprints of the magnetic field lines and layer
parameters\citep{crz00}. The shape of the polar cap could be
determined by the footprints of the last closed field lines
anchored in the stellar surface, and we can label the coordinates
of these footprints as $(x_0, y_0, z_0)$. Then we are able to
define another set of footprints of magnetic field lines
$(x^{'}_0, y^{'}_0, z^{'}_0)$ by multiplying a factor $a_1$ called
layer parameter: $x^{'}_0=a_1x_0$, $y^{'}_0=a_1y_0$, and
$z^{'}_0=\sqrt{1-(x^{'2}_0+y^{'2}_0)}$, where $a_1>1$ indicates the
closed region, and $a_1<1$ represents the open ones.

Since the magnetosphere is co-rotating with the neutron star,
aberration effect occurs along the line of sight. Thus, we have
the aberrated emission direction (in the observer's frame)
$\bold{u}^{'}=(u^{'}_{r}, u^{'}_{\theta}, u^{'}_{\phi})$ in the
expression of the direction $\bold{u}=(u_{r}, u_{\theta},
u_{\phi})$ in the co-rotating frame and the rotational speed
$\beta=|\bold{r}\times \bold{\Omega}|/c$:
\begin{eqnarray}\label{abb}
u^{'}_{r}&=& \frac{u_r\sqrt{1-\beta^2}}{(1+\beta u_{\phi}c)}
\nonumber \\
 u^{'}_{\theta}&=&
\frac{u_{\theta}\sqrt{1-\beta^2}}{(1+\beta u_{\phi}c)}\nonumber \\
u^{'}_{\phi}&=& \frac{u_{\phi}+\beta c}{(1+\beta u_{\phi}c)}.
\end{eqnarray}
Another effect taken into account is the time of flight, which
differs a lot for the photons originated at different sites inside
the magnetosphere. This effect can lead to the phase difference of
the arrival photons comparable to the light curve period. Combing
these two effects, and choosing the rotational axis as the
$z$-axis, we obtain the phase angle $\Phi$ and the polar angle
$\zeta$ given by \citep{ya97}
\begin{eqnarray}\label{dd}
\cos \zeta &=& u^{'}_z/u^{'},\nonumber \\
\Phi &=& -\phi^{'}_{u^{'}}-\bold{r}\cdot \hat{u}^{'}/R_L,
\end{eqnarray}
where $\phi^{'}_{u^{'}}=\arccos (u^{'}_x/u^{'}_{xy})$ (here $(x,y,
z)$ is the cartesian coordinate system) is the azimuthal angle in
the observer's frame. Choosing
\mbox{\boldmath$\Omega$}-\mbox{\boldmath$\mu$} plane to be the
$x$-$z$ plane, $u_{xy}'$ is the length of the projection of
$\hat{u}'$ on the $x$-$y$ plane.

\subsection{Simulated light curve profiles}

 In the following, we adopt in this paper the magnetic coordinates
($\theta_{\ast}, \phi_{\ast}$) to describe the loci of radiation
regions in the magnetosphere. The polar angle $\theta_{\ast}$ is
defined as the angle with respect to the magnetic axis, instead of
the rotational axis. The emitting region is assumed to be in the
shape of a band along the azimuthal direction, with a relatively
smaller longitudinal thickness compared with its azimuthal width.
As the neutron star is rapidly rotating, we set the phase of the
\mbox{\boldmath$\Omega$}-\mbox{\boldmath$\mu$} plane defined by
the rotational axis and magnetic axis as $\phi_{\ast}=0^0$.
Therefore, the transverse extension of the emitting region along
the azimuthal direction could be expressed as $\Delta
\phi_{\ast}$, which is treated as a parameter in our
numerical simulations in this paper. However, there is no
particular parameter for a quantitative longitudinal thickness
$\Delta \theta_{\ast}$ in our simulation, and we represent it by
the layer parameter $a_1$, instead.

\begin{figure}
\epsscale{0.8}
\plotone{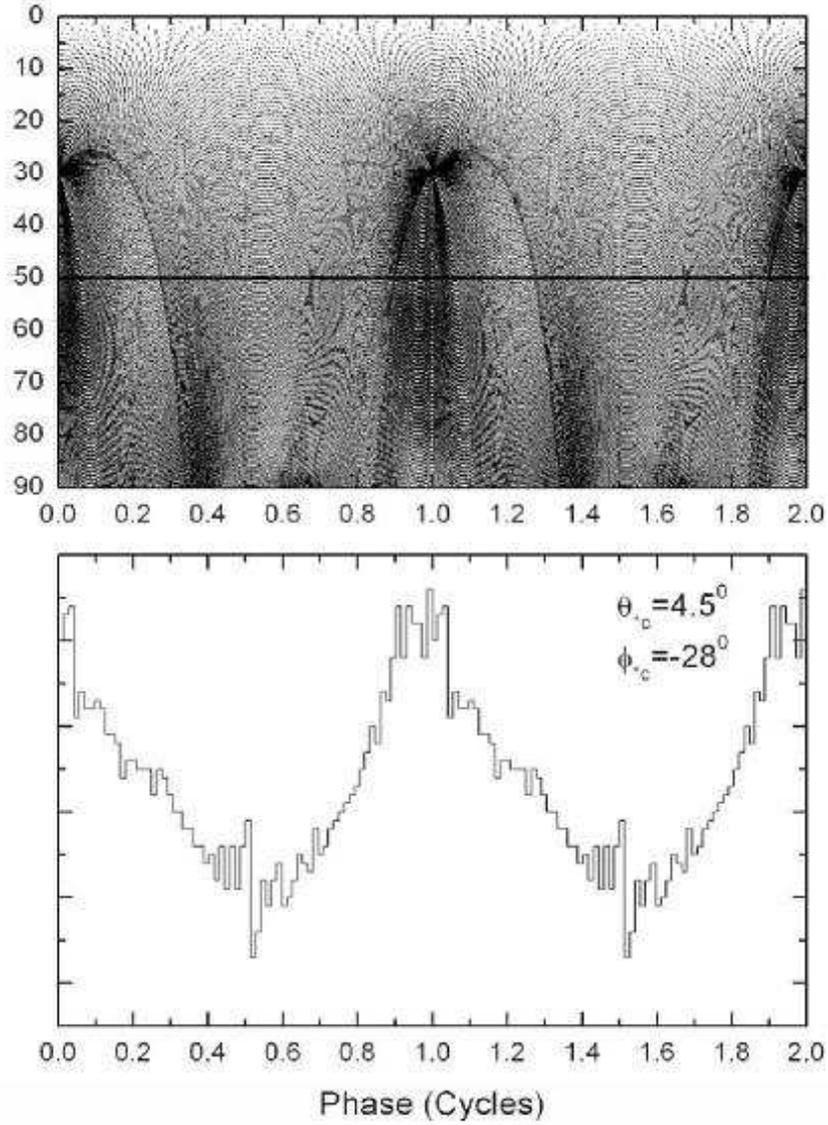}

\caption{Upper: Emission projections on the $(\zeta, \Phi)$-plane
produced by the emitting region with azimuthal width of $\Delta
\phi_{\ast}=180^0$ at layer $a_1=15$. The solid line indicates the
line of sight at the viewing angle $\zeta=50^0$, and the darker
regions correspond to greater intensities. Lower: The single
sinusoidal pattern of pulse profile corresponding to $\zeta=50^0$.
SGR 1860-20 parameters are used in the calculation, and other
parameters for fitting are given in Table.\ref{tab1}.
\label{fig2}}
\end{figure}

\begin{figure}
\epsscale{0.8}
\plotone{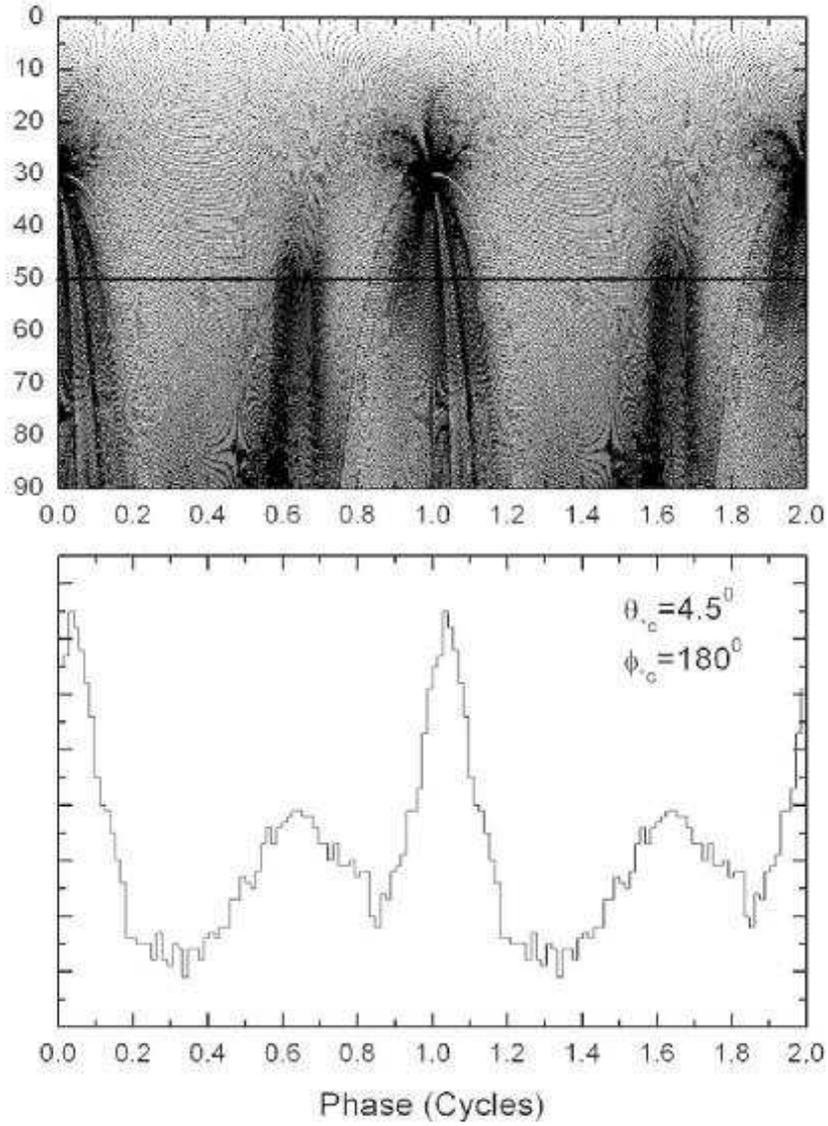}

\caption{Upper: Emission projections on the $(\zeta, \Phi)$-plane.
The same fitting parameter as those in Fig.\ref{fig2} except
$\phi_{\ast c}$. Lower: The double-peaked pattern of pulse profile
corresponding to $\zeta=50^0$.\label{fig3}}
\end{figure}

Combining those factors mentioned above, we show in Fig.\ref{fig2}
and \ref{fig3} two typical light curves generated in the closed
field line zone. Fig.\ref{fig2} is a single sinusoidal pattern,
and Fig.\ref{fig3} is a double-peaked morphology without any
off-pulse phase. The spin period of SGR 1806-20 is applied in the
calculation, and the inclination angle and viewing angle w.r.t.
the spin axis are chosen to be $30^{0}~\rm{and}~50^{0}$. Both
pulse profiles are commonly detected in the SGRs timing
observations (e.g. \cite{woods07}). The azimuthal widths of the
emission regions for these two cases are $\Delta
\phi_{\ast}=180^0$, and the other parameters for these two plots,
like $a_1$ and the ($\theta_{\ast}, \phi_{\ast}$) coordinates of
the emitting regions, are given in Table.\ref{tab1}. In the upper
panels of both figures, we show the emission projections onto the
$(\zeta, \Phi)$-plane to illustrate the two kinds of light curves,
respectively. We can find that the whole $(\zeta, \Phi)$-plane is
fully filled by the emissions from closed field lines, compared
with those partially-filled by the outer gap in the open lines
(e.g. Fig.6-8 in \cite{crz00}). The emission projection of
Fig.\ref{fig3} has a dense bundle in the phase range (0.5, 0.7),
which produces the secondary peak of the double-peaked
light curve.

What makes such difference between these two kinds of
light curves? As shown by \cite{crz00}, the neutron star rotation
results in a non-uniform distribution of the magnetic field lines
in the magnetosphere, i.e., the magnetic field lines are swept
back to accumulate around the
\mbox{\boldmath$\Omega$}-\mbox{\boldmath$\mu$} plane. In our
calculation, the only different parameter in both cases is the
longitude of the emitting region. On the side around
$\phi_{\ast}=0^0$, the accumulated field lines produce the
one-peak dominated light curve profile (e.g. the single sinusoidal
pattern), while on the other side around $\phi_{\ast}=180^0$, the
widely-separated field lines make the double-peaked profile
possible.

\section{Application to SGR 1806-20 in December 2004\label{app}}

\subsection{Pulse profile evolution}

As illustrated in the standard magnetar model, the crust of the
magnetar breaks when the magnetostatic equilibrium in the lower
crust is no longer be sustained, and launches a hot fireball,
which triggers the outburst of SGRs. The released energy comes
from the reconnection of magnetic field lines in a crustal plate,
which can be modelled as
\begin{equation}\label{eq1}
E=\frac{B^2}{8\pi}l^3,
\end{equation}
where $B$ is the magnetic field in the crust, and $l$ is the size
of the crustal plate. The energy released by SGR 1806-20 in the
Dec 27 2004 burst is estimated to be as high as $\sim 10^{46}$erg,
and by substituting the inferred magnetic field strength of order
$10^{15}~\rm{G}$, we can estimate $l\sim 10^5~\rm{cm}$, which is
about the thickness of the neutron star crust. A clump of
electron-positron or electron-proton plasma is then ejected into
the magnetosphere, and trapped by the magnetic field lines
anchored in such crustal platelet. The charged particles emitted
by the hot plasma travel along the closed magnetic field lines,
and radiate photons.   In the following emission morphology simulation, we
will assume the emissivity along the field lines is uniform. However, the
emission region along the azimuthal and polar directions have
a finite characteristic dimension corresponding to $\sim 10^5$cm. We assumed that either 
the crustal motion
driven by the neutron vortex \citep{ruderman91} or the lateral
motion of the plasma across the field lines driven by the residual
electric field, could lead to the change of the radiation
morphologies. We will discuss which mechanism is more plausible
later.

\begin{figure}
\epsscale{0.8}
\plotone{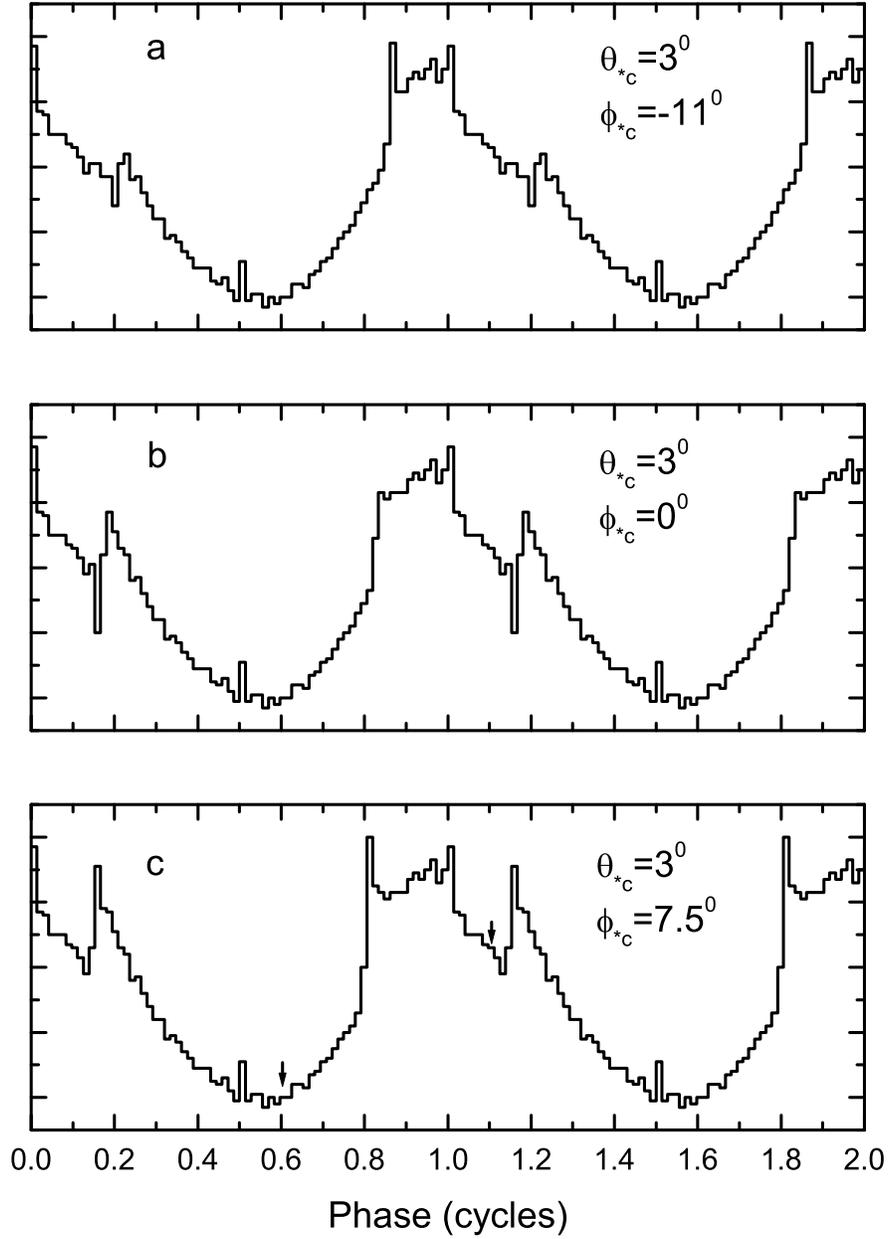}

\caption{Simulated pulsed profiles of SGR 1806-20 in the giant
flare on 27 December 2004. The emitting region drifts along the
azimuthal direction, with the center coordinates indicated. The
arrows in panel c indicate the positions of the minor peaks of the
radiation from both magnetic poles.}\label{fig4}
\end{figure}

We attempted to simulate the pulse profile evolution during the
giant flare of 27 December 2004 (e.g. Figure 2 in
\cite{palmer05}). The most significant feature of the SGR 1806-20
pulsed radiation is the increased amplitude of the secondary peak
related to the primary one. Our trials on the numerical simulation
suggest that such change is due to the motion of emitting region.
We present our results in Fig.\ref{fig4}, which shows the effect
of the azimuthal motion of the emitting region across the magnetic
field lines with the typical layer parameter $a_1=10$, anchored in
the crustal plate with the size about $1.5\times 10^5$ cm.
Fig.\ref{fig4} also shows some other features (e.g. the width of
the peaks and the phase separation between two peaks) to be
consistent with the observation. The inclination angle of the
magnetic dipole and the viewing angle of the observer are chosen
to be $30^0$ and $50^0$, respectively. We assume that size of the
plasma or the emitting region remains unchanged during the motion,
and we describe the loci of the emitting region center in terms of
the $(\theta_{\ast c},\phi_{\ast c})$, which are the magnetic
coordinates of the middle point of the emitting region at the
typical layer. Since the motion is along the azimuthal direction,
$\theta_{\ast c}=a_1\times \theta_p=3^0$ for all three panels in
Fig.\ref{fig4}. As the
\mbox{\boldmath$\Omega$}-\mbox{\boldmath$\mu$} plane is defined as
$\phi_{\ast}=0^0$, the azimuthal angle $\phi_{\ast}$ increases
along the spin direction of the neutron star. The center of the
radiation region was shifted from $\phi_{\ast c}=-11^0$ (panel a
in Fig.\ref{fig4}) to $\phi_{\ast c}=0^0$ (panel b in
Fig.\ref{fig4}), and finally at $\phi_{\ast c}=7.5^0$ (panel c in
Fig.\ref{fig4}) All parameters applied in our model
fitting are listed in Table.\ref{tab1}.

As shown by \cite{arendt98}, the polar cap of a rotating
neutron star is asymmetrical and even probably discontinuous,
which could make the radiation morphology complicated and
asymmetrical. In our simulation, the primary peak at phase 0.9 of
the light curve results from the majority of the magnetic field
lines in which the plasma is trapped, and only a small portion of
the field lines on the right edge of the plasma generate the
secondary peak (at phase 1.2), where 'right' means the site whose
$\phi_{\ast}$ value is larger. As the plasma is drifted from left
to right (along the direction $\phi_{\ast}$ increases), more
magnetic field lines at the right site are enrolled to radiate
photons, which makes the secondary peak grow up. The radial
distance of the local place where the arrival photons are
generated is shown in Fig.\ref{fig6}. Since the magnetosphere
co-rotates with the neutron star, we calculate the modelling
velocity of the emitting region motion in $\phi_{\ast}$ direction.
As shown in Fig.\ref{fig7}, the speed of the drift is estimated to
be $\sim 10^4~\rm{cm~s^{-1}}$ according to the pulse profile
evolution timing by observation. We also give the expected
radiation from the two magnetic poles in panel c of Fig.
\ref{fig4}, which are indicated by two arrows. As the definition
of the $(\Omega, \mu)$-plane, the magnetic poles are located at
$\phi_{\ast}=0~\rm{and}~\pi$ (0 and 0.5 in the phase cycle),
respectively. However, the time of flight effect makes the
radiation from magnetic poles has a tiny shift in phase, e.g. 0.1
and 0.6. At the early stage, the flux of the radiation from the
trapped plasma is so strong that the emissions from the polar caps
could not be resolved. However, as the intensity of the two main
peaks fades, the two minor peaks might become
discriminable.

\begin{figure}
\epsscale{0.8}
\plotone{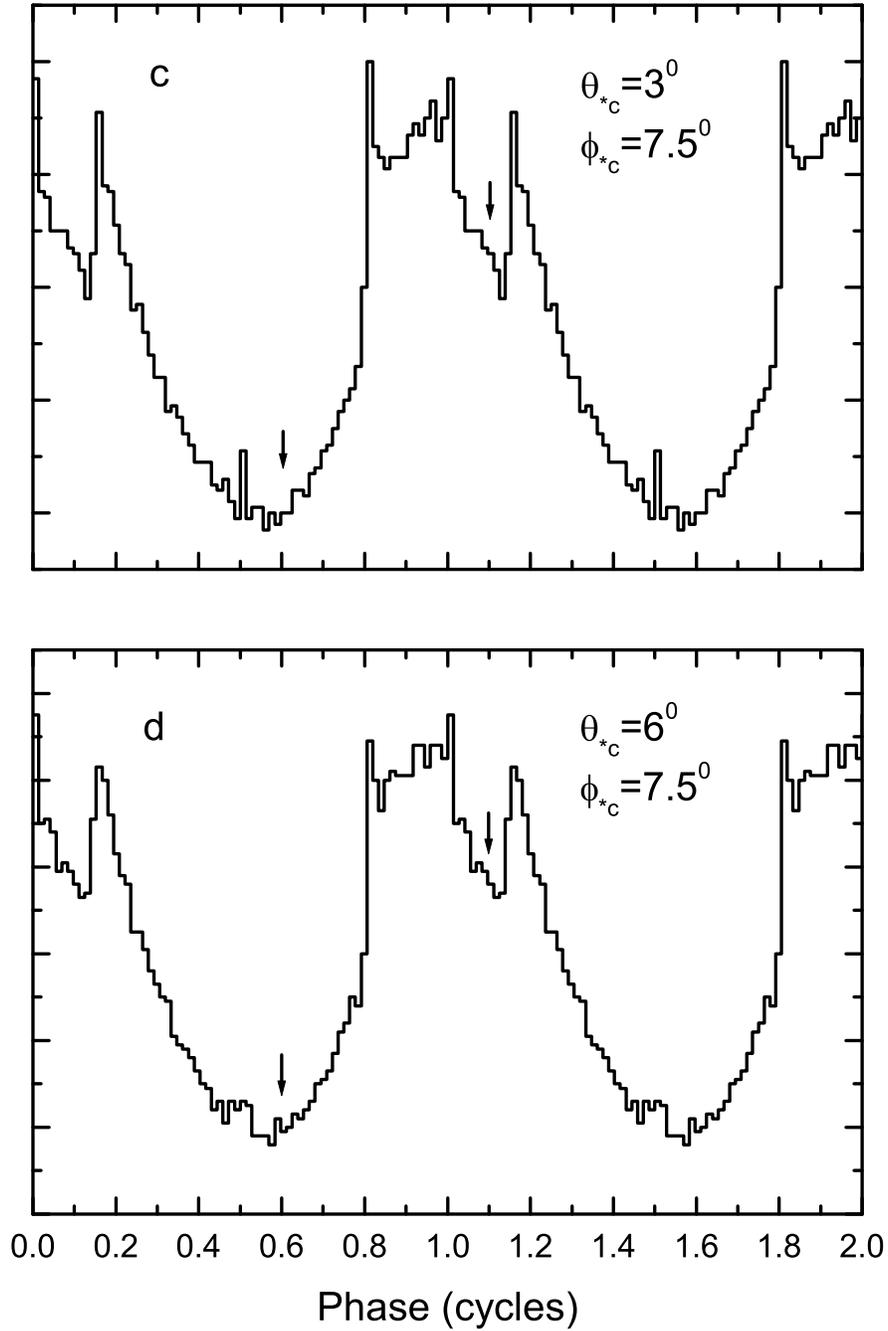}

\caption{Simulated pulsed profiles of SGR 1806-20 in the giant
flare on 27 December 2004. The arrows indicate the positions of
the minor peaks of the radiation from both magnetic poles. The
emitting region drifts in the direction perpendicular to that in
Fig.\ref{fig4}.}\label{fig5}
\end{figure}

\begin{figure}
\epsscale{0.8}
\plotone{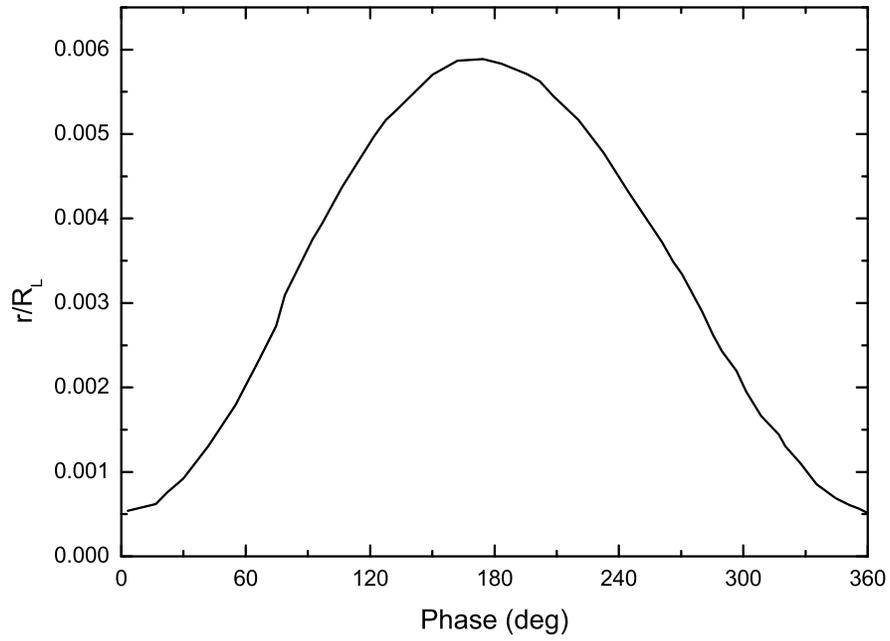}

\caption{Variation of radial distance of the local position where
the arrival photons are generated with the pulse phase in
Fig.\ref{fig4}.}\label{fig6}
\end{figure}

\begin{figure}
\epsscale{0.8}
\plotone{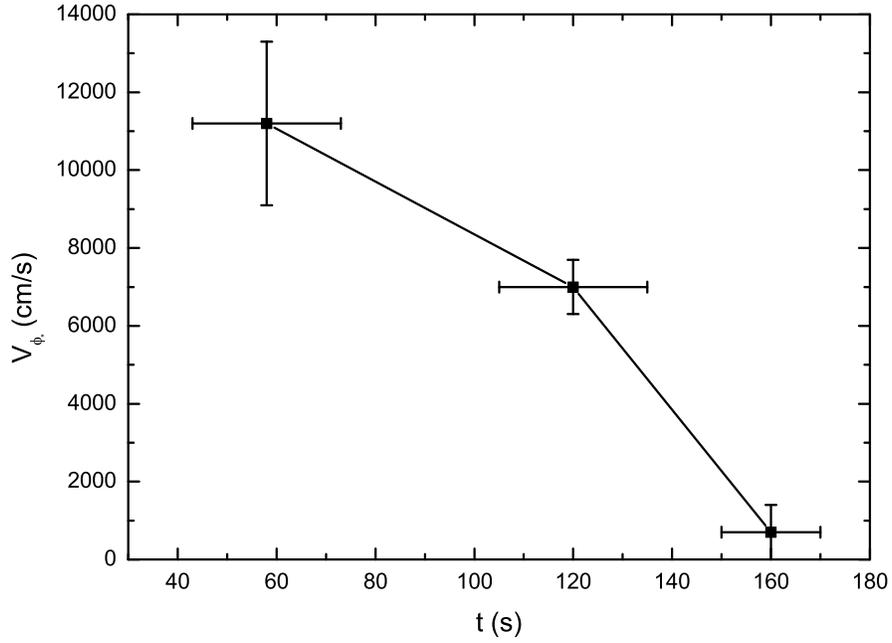}

\caption{The speed of the emitting region drift in $\phi_{\ast}$
direction is calculated by the comparison between our numerical
results in Fig.\ref{fig4} and the observation. The speed is given
by $v_{\phi_{\ast}}(t=\frac{t_1+t_2}{2})=\Delta x/\Delta t$, where
$\Delta x$ is the displacement of the emitting region centers in
two adjacent panels in Fig.\ref{fig4}, and $\Delta t=t_2-t_1$ is
the time interval. $\Delta t$ and error of $t$ is roughly
determined from the time scale of Supplementary Figure 1 in
\cite{palmer05}. The error of $v_{\phi_{\ast}}$ is calculated from
the error propagation $\sigma
(v_{\phi_{\ast}})/\bar{v}_{\phi_{\ast}}=\sigma(t)/\bar{t}$. Since
the emitting region changes the direction of movement or stops
moving after $t=160~\rm{s}$, we set
$v_{\phi_{\ast}}(t=160~\rm{s})\approx 0$.}\label{fig7}
\end{figure}

In general, the emitting region should not always move in one
direction. Fig. \ref{fig5} shows the pulse profile evolution when
the motion of the radiation region is changed to be along the
$\theta$-direction. Panel c is the same as the one in Fig.
\ref{fig4}, and the light curve of panel d is produced when the
center of plasma is located at $a_1$=20, with the same azimuthal
position as that in panel c. We find that the morphology of the
pulsed emission remains roughly unchanged, corresponds to the late
stage ($\sim$ 170s after the burst) of tail in the giant flare of
27 December 2004. We want to remark that the light curve can also
remain unchanged when the motion of the radiation region stops. We
cannot differentiate these two possibilities from the light curve
evolution.

\subsection{Interpretation of numerical results}

In order to simulate the time evolution of the light curves
of SGR 1806-20, we have assumed that the emitting region can migrate
in the magnetosphere. In the following, we would
like to discuss several possible movements induced in the
crust and the magnetosphere, and their resultant speeds.

The interaction between
the flux tubes and vortex lines in the core of neutron star make
them inter-pin to each other. The vortex lines will move out of the
core due to the spin-down of the star, they drag the flux tubes with them.
However,  the flux tubes are anchored in the crust, consequently a large stress
will be applied to the crust from the flux tubes (e.g. Ruderman 1991). When the
crust breaks, the flux tubes will try to reduce
their tension and drag the broken crust platelet to move.
The maximum shear
stress on the base of the crust from core magnetic flux tube
motion which the crust could sustain is
\begin{equation}\label{smax}
S\sim \frac{BB_{crit}}{8\pi}\sim 3\times 10^{29}~\rm{dyn~cm^{-2}},
\end{equation}
where $B_{crit}\sim10^{16}~\rm{G}$ is the critical magnetic field inside the
magnetic flux tube. However, the crust may already break before
the magnetic stress reaches the maximum value. The shear modulus
$\mu$ of a 1 km thick crust could not be larger than
$10^{30}~\rm{dyn~cm^{-2}}$, and the crust may break when the
stress reaches
\begin{equation}\label{fstress}
S_{break}\sim f\mu\theta_{max},
\end{equation}
where $f$ is a factor of order unity, and $\theta_{max}$ is the
yield strain under tension or compression, which is
$\sim 10^{-1}-10^{-3}$ (c.f. Ruderman 1991).

Another mechanism to drive the crust to move is the vortex
creeping. As the rotation of the neutron star slows down, the flux
tubes are driven outwardly by the the neutron vortices. The force
acting on unit length of a flux tube at the core-crustal interface
is \citep{chau92,ding93}
\begin{equation}\label{fn}
f_n=\frac{2\Phi_{0}\rho_cR_c\Omega_s\omega_{cr}}{B_c},
\end{equation}
where $\Phi_0=hc/2e\sim 2\times 10^{-7}\rm{G/cm^2}$ is the flux
quantum, and the subscript $c$ represents the values in the core.
$\Omega_s$ is the rotation rate of the core superfluid, and can be
approximated as that of the crust $\Omega_c$ or the spin rate of
the star in our following estimation. $\omega_{cr}$ is the maximum
angular velocity lag between $\Omega_s$ and $\Omega_c$. $B_c$ is
the core magnetic field, and could be only about $10^{-3}$ of the
surface value because the flux tubes are pushed out of the core due to
spin-down (Ding, Cheng \& Chau 1993).
Thus, we can estimate the total magnitude of the force acting on
the crust platelet with area $A$. The number of flux tubes anchored in this
platelet is given by
\begin{eqnarray}\label{nf}
N_f &=& N_{total}\frac{A}{\pi R^2_c} \nonumber \\
    &=& \frac{A}{\pi R^2_c}\frac{\pi R^2_c
    B_c}{\Phi_0}=\frac{AB_c}{\Phi_0},
\end{eqnarray}
Substituting $f_n$ in Eq.\ref{fn}, we have the total driving force on a platelet
\begin{equation}\label{bigfn}
F_n=N_ff_nR_c=2A\rho_cR^2_c\Omega_s\omega_{cr}.
\end{equation}
We then make a dimension analysis to estimate the velocity
\begin{equation}\label{vcreep}
<v>\sim \sqrt{F/\rho A}=(R^2_c\Omega_s\omega_{cr})^{1/2}.
\end{equation}
$\omega_{cr}$ is given by various models as of order
$10^{-6}~\rm{rad~s^{-1}}$ \citep{chau92}, therefore, we have the
velocity of the flux tube about $10^3~\rm{cm~s^{-1}}$.

However, when the flux tubes move, they will experience a drag force by
the electron sea in the core. The drag force per unit length of a single flux
tube is given by
\begin{equation}\label{fv}
f_v(v_p)=\frac{3\pi}{64}\frac{n_ee^2\Phi^2_0v_p}{E_f\Lambda c}.
\end{equation}
Here, $n_e\sim 10^{37}~\rm{cm^{-3}}$ is the electron density, and
$E_f$ is the electron Fermi energy, which is about 100 MeV.
The penetration length of a proton $\Lambda$ is $\sim
100~\rm{fm}$, and $v_p$ the velocity of the flux tube. By equating
the vortex acting force $f_n$ and drag force $f_v$, we calculate
the velocity as
\begin{eqnarray}\label{vp}
v_p &=& \frac{64}{3\pi}\frac{E_f\Lambda
c\rho_cR_c\Omega_s\omega_{cr}}{B_cn_ee^2\Phi_0}
\nonumber \\
    &\approx& 10^{-12} ~\rm{cm~s^{-1}},
\end{eqnarray}
and \cite{ruderman98} have considered the collective motion of
flux tubes and gave an estimation of $10^{-6} ~\rm{cm~s^{-1}}$. If
we equate the magnetic stress force $SA$ and the drag force by
\begin{eqnarray}\label{dd}
\frac{BB_c}{8\pi}A=f_v(v_p)N_fR_c,
\end{eqnarray}
we have the velocity
\begin{equation}\label{vp2}
v_p=\frac{64}{24\pi^2}\frac{BE_f\Lambda c}{n_ee^2\Phi_0R_c}
\end{equation}
at about $10^{-7}\sim10^{-8}~\rm{cm~s^{-1}}$. Therefore, we find
the velocity of the crustal motion is too small to account for the
emitting region drift.

On the other hand, if we consider the drift velocity of the plasma driven by
the electric field in the magnetosphere, we can write
\begin{equation}\label{vd}
v_d(D)\sim \frac{\delta \bold{E}\times \bold{B}}{B^2}c\sim \left |
\frac{\delta E}{B(D)} \right | c,
\end{equation}
where $\delta E$ is the residual electric field which drives the
plasma to move, and $B(D)$ is the local magnetic field at the
radial distance $D$, where the emitting plasma is located. As
shown in Fig.\ref{fig6}, the average radial distance of the
emitting region is about $D\sim 0.002R_L=0.7\times 10^8~\rm{cm}$.
The length of the magnetic flux loop at $a_1=10$ is of order about
$\mathcal{L}\sim 0.01R_L$ referred to Fig.\ref{fig1}. The residual
electric field could be determined by $\delta E\approx \Delta
V/\mathcal{L}$, where $\Delta V$ is the electric potential drop.

In closed field it is not clear how any substantial potential can
survive because electrons and positrons can be created and trapped
in the closed field line region. These electron/positron pairs can
screened the electric field. In our simulation, we find that the
emission region is characterized by $a_1$=10, which is not far
away from the open field line region. For a platelet with
characteristic dimension of $\sim 10^5$cm, we can imagine that
part of the platelet is the open field line region, where a
characteristic potential $\Delta V\approx 6.6\times
10^{15}B_{15}/P^2~\rm{volts}$ can be maintained (Ruderman \&
Sutherland 1995). Here $B_{15}$ is the surface magnetic field in
unit of $10^{15}~\rm{G}$. Thus, we have $\delta E\sim 1\times
10^{3}B_{15}~\rm{esu~cm^{-2}}$. Therefore, Eq.\ref{vd} could be
re-written as
\begin{eqnarray}\label{vd2}
v_d &\approx& \frac{\delta E}{B_{s}(R_{\ast}/D)^3}c \nonumber \\
    &\approx& 3\times 10^{4}D^3_{8}~\rm{cm~s^{-1}},
\end{eqnarray}
where $D_8$ is the radial distance of the emitting region in unit
of $10^8~\rm{cm}$. By substituting the average value $D_8=0.7$,
the drift velocity of the emitting region is
$v_d=10^4~\rm{cm~s^{-1}}$, which is consistent with our result in
Fig.\ref{fig7}. Furthermore, as illustrated in Eq.\ref{vd}, the
drift velocity is proportional to the magnitude of the residual
electric field, we may also argue that when $\delta E$ decays as
the equilibrium charge distribution in the open field line region
is being re-established, the drift velocity becomes smaller. Even
the emitting region could stop drifting and finally stays in the
location indicated in panel c of Fig.\ref{fig4}, where it emits
the rest radiation. If such case applies, it becomes unnecessary
for us to propose the sudden change of motion direction in
Fig.\ref{fig5}.

\section{Summary and Discussion \label{dis}}

We have investigated the timing properties of SGRs in the
persistent state and outburst tails, and ascribed the variety of
the pulsed radiation morphologies as the emissions coming from the
closed field line regions inside the neutron star magnetosphere.
For the weak relativistic effects in the magnetosphere closer to
the stellar surface, the totally different features of the light
curve of SGRs from those of Crab and Vela could be explained.
Furthermore, by assuming the emitting region drift in the
magnetosphere, we are able to simulate the pulse profile evolution
of SGR 1806-20 during the giant flare on 27 December 2004.
In addition, when we take the emissions from both magnetic poles
into account, we are also able to explain the occurrence of the
third peak with the phase roughly consistent with the observation,
which increases in strength relative to the two major
ones (e.g. Figure 2 in \cite{palmer05}) at the end of the tails after
the outburst. We believe that when the emission mechanism becomes clear, the
emissivity on the field lines should not be uniform. In that case we may be
able to predict a stronger third peak.

\cite{feroci01} made simple analytic fits to the flare light
curves of SGR 1900+14 on 27 August 1998, and concluded that a
fireball with contracting surface (rather than a cooling surface
of fixed area) could provide a reasonable explanation to the decay
tail of the outburst. They also proposed that the four-peaked
profile was produced by several X-ray jets tied to the neutron
star surface. However, the size of the radiation region in our
fitting of SGR 1806-20 is unchanging, and the smooth decay in the
tail phase of the giant flare might be due to the cooling of the
plasma. It was believed that the phase stability of the pulses in
light curve is due to the fixed location of the emitting region on
the stellar surface. But our results show that a small migration
could not break the stability in phase.

We have speculated some possible mechanisms, which cause the radiation region to
migrate. It seems very clear that the physical motion of magnetic field lines,
where the charged particles are trapped, must be very slow due to the extremely
large drag force by the electrons in the core of neutron star. On the other hand,
the $\bf{E}\times \bf{B}$ drift seems more possible. However, how this residual
electric field survives from the screen of electron/positron pairs is not clear.
We argue that one possible way is that some platelet is in the open field lines.
It is unclear if this situation always occurs.

The 3-100 keV phase-averaged spectrum of the pulsed tail during
the 2004 burst is fitted by a blackbody function at the
temperature of 5.1 keV plus a power law \citep{hurley05}. We
didn't give the calculated spectrum in this paper, for the
radiation mechanism in the closed field lines needs further work.
However, we argue that the power law component results from the
synchrotron radiation by the charged particles gyrating along the
magnetic field lines. We need more information in the higher
energy band and the phase-resolved spectra to provide more
constrains and modification on our geometrical model.

\acknowledgments

We thank C. Thompson and M. Ruderman for useful discussions and the anonymous
referee for helpful comments to improve the paper. This
work is supported by a RGC grant of Hong Kong SAR Government, and
Y. F. Huang is also supported by the National Natural Science
Foundation of China (Grants 10625313 and 10221001).

\clearpage

\clearpage



\begin{deluxetable}{lrrrrcrrrrr}

 \tablewidth{0pt} \tablecaption{Parameter sets in
Fig.\ref{fig2}-\ref{fig5}.}
\tablehead{ \colhead{plot label} &
\colhead{typical layer $a_1$} & \colhead{$\theta_{\ast c} (^0)$} &
\colhead{$\phi_{\ast c} (^0)$} & \colhead{$\Delta \Phi_{\ast} (^0)$}}
 \startdata
  Fig.\ref{fig2} & 15~~~~~~& 4.5 & -28 ~& $180$ \\
  Fig.\ref{fig3} & 15~~~~~~& 4.5 & 180 ~& $180$ \\
  panel a in Fig.\ref{fig4} & 10~~~~~~& 3.0 & -11 ~& $180$ \\
  panel b in Fig.\ref{fig4} & 10~~~~~~& 3.0 & 0 ~& $180$ \\
  panel c in Fig.\ref{fig4} & 10~~~~~~& 3.0 & 7.5 ~& $180$ \\
  panel d in Fig.\ref{fig5} & 20~~~~~~& 6.0 & 7.5 ~& $180$ \\
\enddata
\tablecomments{$(\theta_{\ast c},~\phi_{\ast c})$ are the magnetic
coordinates of the middle point of the emitting region, where
$\theta_{\ast c}=a_1\times \theta_p$ is the product of the layer
parameter and the angular radius of the polar cap, and $\phi_{\ast
c}=0^0$ is defined as the
\mbox{\boldmath$\Omega$}-\mbox{\boldmath$\mu$} plane.
$\Delta \phi_{\ast}$ is the transverse extension of the emitting
region along the azimuthal direction.
The magnetic inclination angle is $\alpha=30^0$, and the viewing
angel is $50^0$.}\label{tab1}

\end{deluxetable}



\clearpage

\begin{thebibliography}{}

\bibitem[Arendt \& Eilek (1998)]{arendt98} Arendt, P. N., \&
Eilek, J. A. 1998, astro-ph/9801257

\bibitem[Chau, Cheng \& Ding (1992)]{chau92} Chau, H. F., Cheng,
K. S., \& Ding, K. Y. 1992, \apj, 399, 213

\bibitem[Cheng, Ho \& Ruderman (1986)]{chr86} Cheng, K. S., Ho,
C., \& Ruderman, M. A. 1986, \apj, 300, 500

\bibitem[Cheng, Ruderman \& Zhang (2000)]{crz00} Cheng, K. S.,
Ruderman, M. A. \& Zhang, L. 2000, \apj, 537, 964

\bibitem[Chiang \& Romani (1994)]{cr94} Chiang, J., \& Romani, R.
W. 1994, \apj, 436, 754

\bibitem[Daugherty \& Harding (1996)]{dh96} Daugherty, J. K., \&
Harding, A. K. 1996, \apj, 458, 278

\bibitem[Ding, Cheng \& Chau (1993)]{ding93} Ding, K. Y., Cheng,
K. S., \& Chau, H. F. 1993, \apj, 408, 167

\bibitem[Dyks \& Rudak (2003)]{dr03} Dyks, J., \& Rudak, B. 2003,
\apj, 598, 1201

\bibitem[Fern\'{a}ndez \& Thompson (2007)]{thompson07} Fern\'{a}ndez, R., \&
Thompson, C. 2007, \apj, 660, 615

\bibitem[Feroci et al. (2001)]{feroci01} Feroci, M., Hurley, K., Duncan, R. C.
\& Thompson, C. 2001, \apj, 549, 1021

\bibitem[Goldreich \& Julian (1969)]{gj69} Goldreich, P., \&
Julian, W. H. 1969, \apj, 157, 869

\bibitem[Harding (1981)]{harding81} Harding, A. K. 1981, \apj,
245, 267

\bibitem[Hurley et al. (1999)]{hurley99}Hurley, K., et al. 1999, Nature, 397,
 41

\bibitem[Hurley et al. (2005)]{hurley05}Hurley, K., et al. 2005,
Nature, 434, 1098

\bibitem[Jia et al. (2007)]{jia07}Jia, J. J., Tang, A. P. S.,
Takata, J., Chang, H. K., \& Cheng, K. S. 2007, J. Adv. Space
Res., in press

\bibitem[Mazets et al. (1979)]{mazets79} Mazets, E. P., et al. 1979,
Nature, 282, 587

\bibitem[Palmer et al. (2005)]{palmer05} Palmer, D. M., et al.
2005, Nature, 434, 1107

\bibitem[Romani \& Yadigaroglu (1995)]{romani95} Romani, R. W., \&
Yadigaroglu, I.-A. 1995, \apj, 438, 314

\bibitem[Ruderman (1991)]{ruderman91} Ruderman, M. 1991, \apj,
382, 587

\bibitem[Ruderman \& Sutherland (1975)]{ruderman75} Ruderman, M.,
\& Sutherland, P. A. 1975, \apj, 196, 51

\bibitem[Ruderman, Zhu \& Chen (1998)]{ruderman98} Ruderman, M.,
Zhu, T. H., \& Chen, K. Y. 1998, \apj, 492, 267

\bibitem[Thompson \& Duncan (1995)]{td95} Thompson, C., \& Duncan,
R. C. 1995, \mnras, 275, 255

\bibitem[Thompson \& Duncan (1996)]{td96} Thompson, C., \& Duncan,
R. C. 1996, \apj, 473, 322

\bibitem[Thompson, Lyutikov \& Kulkarni (2002)]{tlk02} Thompson,
C., Lyutikov, M., \& Kulkarni, S. R. 2002, \apj, 574, 332

\bibitem[Woods et al. (2007)]{woods07} Woods, P. M., et al. 2007,
\apj, 654, 470

\bibitem[Woods \& Thompson (2004)]{woods04} Woods, P. M., \&
Thompson, C. 2004, astro-ph/0406133

\bibitem[Yadigaroglu (1997)]{ya97} Yadigaroglu, I.-A. 1997, Ph.D.
thesis, Stanford Univ.

\end{thebibliography}
\end{document}